\documentclass[a4paper,10pt]{article}
\usepackage{setspace}
\usepackage{longtable}
\usepackage{fullpage}
\usepackage[latin1]{inputenc}
\usepackage[left]{eurosym}
\usepackage{amsmath,amsthm,amssymb,amstext,array,rotating,multicol,theorem}

\usepackage{psfrag}
\usepackage{subfigure}

\usepackage{pstricks}
\usepackage{epsfig}
\usepackage{pst-grad} 
\usepackage{pst-plot} 

\usepackage{setspace} 		

\newcommand{\nn}{\nonumber \\}
\newcommand{\lucin}[1]{#1}
\newcommand{\lucout}[1]{}
\DeclareMathOperator{\D}{d\!}
 \DeclareMathOperator{\ee}{e}

\title{L2 OSTC-CPM: Theory and design}
\author{Matthias Hesse\thanks{The work of Matthias Hesse is supported by a EU Marie-Curie Fellowship
(EST-SIGNAL program : http://est-signal.i3s.unice.fr) under contract No
MEST-CT-2005-021175.}, J\'er\^ome Lebrun and Luc Deneire\\Lab. I3S, CNRS, University of Nice, Sophia Antipolis, France\\{E-Mail:\tt \{hesse,lebrun,deneire\}@i3s.unice.fr}}
\date{}
\begin{document}
\maketitle

The combination of space-time coding (STC) and continuous phase
modulation (CPM) is an attractive field of research because both STC
and CPM bring many advantages for wireless communications. 
Zhang and Fitz \cite{Zhan00} were the
first to apply this idea by constructing a trellis based scheme. But
for these codes the decoding effort grows exponentially with the
number of transmitting antennas. This was circumvented by orthogonal
codes introduced by Wang and Xia \cite{Wang05}. Unfortunately, based
on Alamouti code \cite{Alam98}, this design is restricted to two
antennas.

However, by relaxing the orthogonality condition, we prove here that
it is possible to design $L^2$-orthogonal space-time codes which
achieve full rate and full diversity with low decoding effort. In part
one, we generalize the two-antenna code proposed by Wang and Xia
\cite{Wang05} from pointwise to $L^2$-orthogonality and in part two we
present the first $L^2$-orthogonal code for CPM with three antennas.
In this report, we detail these results and focus on the properties of
these codes. Of special interest is the optimization of the bit error
rate which depends on the initial phase of the system. Our simulation
results illustrate the systemic behavior of these conditions.

\section{Part one: Two antennas case}
To combine the high power efficiency of Continuous Phase Modulation
(CPM) with either high spectral efficiency or enhanced performance in
low Signal to Noise conditions, some authors have proposed to
introduce CPM in a MIMO frame, by using Space Time Codes (STC).  In
this part, we address the code design problem of Space Time Block
Codes combined with CPM and introduce a new design criterion based on
$L^2$ orthogonality.  This $L^2$ orthogonality condition, with the
help of simplifying assumption, leads, in the 2x2 case, to a new
family of codes.  These codes generalize the Wang and Xia code, which
was based on pointwise orthogonality.  Simulations indicate that the
new codes achieve full diversity and a slightly better coding gain.
Moreover, one of the codes can be interpreted as two antennas fed by
two conventional CPMs using the same data but with different alphabet
sets.  Inspection of these alphabet sets lead also to a simple
explanation of the (small) spectrum broadening of Space Time Coded
CPM.

\subsection{Introduction}
Since the pioneer work of Alamouti \cite{Alam98} and Tarokh
\cite{Taro99a}, Space Time Coding  has been a fast growing field
of research where numerous coding schemes have been introduced.
Several years later Zhang and Fitz \cite{Zhan00,Zhan03} were the first
to apply the idea of STC to continuous phase modulation (CPM) by
constructing trellis codes. In \cite{Zaji07} Zaji\'c and St\"uber
derived conditions for partial response STC-CPM to get full diversity
and optimal coding gain.  A STC for noncoherent detection based on
diagonal blocks was introduced by Silvester et al. \cite{Silv06}.

The first orthogonal STC for CPM for full and partial response was
developed by Wang and Xia \cite{Wang04,Wang05}. The scope of this
part is also the design of an orthogonal STC for CPM. But unlike
Wang-Xia aprroach \cite{Wang05} which starts from a QAM orthogonal
Space-Time Code (e.g. Alamouti's scheme \cite{Alam98}) and modify it
to achieve continuous phases for the transmitted signals, we show here
that a more general $L^2$ condition is sufficient to ensure fast
maximum likelihood decoding with full diversity.

\begin{figure}
  \centering \includegraphics[width=0.5\textwidth]{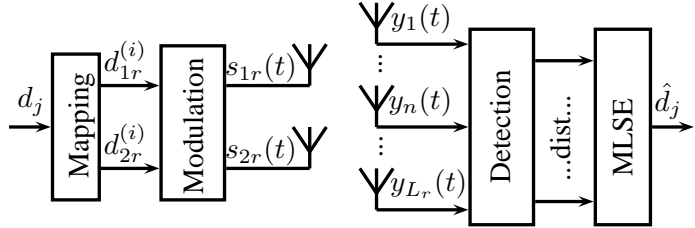}
  \caption{Structure of a MIMO Tx/Rx system }
  \label{fig:block1}
\end{figure}

In the considered system model (Fig.~\ref{fig:block1}), the data
sequence $d_j$ is defined over the signal constellation set
\begin{equation}
  \Omega_{d}=\{ -M+1,-M+3,\ldots,M-3,M-1\}
\end{equation}
for an alphabet with $\log_2M$ bits. To obtain the structure for a
\lucin{Space Time} Block Code (\lucin{ST}BC) this sequence is mapped
to data matrices $\mathbf D^{(i)}$ with elements $d_{mr}^{(i)}$, where
$m$ denotes the transmitting antenna, $r$ the time slot into a block and
$(i)$ a parameter for partial response CPM. The data matrices are then
used to modulate the sending matrix
\begin{equation}
  \label{eq:symmatrix}
  {\bf S}(t) = \begin{bmatrix}
    s_{11}(t) & s_{12}(t) \\
    s_{21}(t) & s_{22}(t)
  \end{bmatrix}\text{.}
\end{equation}
Each element is defined for $(2l+r-1)T\leq t\leq(2l+r)T$ as
\begin{equation}
  s_{mr}(t) = \sqrt{\frac{E_s}{T}}\ee^{j2\pi\phi_{mr}(t)} 
\end{equation}
where $E_s $ is the symbol energy and $T$ the symbol time. The phase
$\phi_{mr}(t)$ is defined in the conventional CPM manner \cite{Ande86}
with an additional  correction factor $c_{mr}(t)$ and is
therewith given by
\begin{equation}
  \phi_{mr}(t)=\theta_m(2l+r) + 
  h\!\!\!\!\!\!\!\!\!\sum\limits_{i=2l+1+r-\gamma}^{2l+r}
  \!\!\!\!\!\!\!\!\!d_{mr}^{(i)}q(t-(i-1)T)+c_{mr}(t)
  \label{eq:phase}
\end{equation}
where $h=2m_0/p$ with $m_0$ and $p$ relative primes is called the
modulation index. The phase smoothing function $q(t)$ has to be a
continuous function with $q(t)=0$ for $t\leq0$ and $q(t)=1/2$ for
$t\geq \gamma T$.

The memory length $\gamma$ determines the length of $q(t)$ and affects
the spectral compactness. For large $\gamma$ we obtain a compact
spectrum but also a higher number of possible phase states which
increases the decoding effort. For full response CPM, we have
$\gamma=1$ and for partial response systems $\gamma>1$.

 The choice of the correction factor $c_{mr}(t)$ in Eq.
(\ref{eq:phase}) is along with the mapping of $d_j$ to $\mathbf
D^{(i)}$, the key element in the design of our coding scheme. It will
be detailed in Section \ref{sec:Cond}. We then define $\theta_m(2l+r)$
in a most general way
\begin{equation}
  \theta_m(2l+3)=\theta_m(2l+2)+\xi(2l+2) 
  =\theta_m(2l+1)+\xi(2l+1)+\xi(2l+2).
  \label{eq:theta1}
\end{equation}
The function $\xi(2l+r)$ will be fully defined from the contribution
$c_{mr}(t)$ to the phase memory $\theta_m(2l+r)$. For conventional CPM
system, $c_{mr}(t)=0$ and we have
$\xi(2l+1)=\frac{h}{2}d_{2l+1-\gamma}$.

 The channel coefficients $\alpha_{mn}$ are assumed to be
Rayleigh distributed and independent. Each coefficient $\alpha_{mn}$
characterizes the fading between the $\mathrm{m^{th}}$ transmit (Tx) antenna and
the $\mathrm{n^{th}}$ receive (Rx) antenna where $n=1,2,\ldots,L_r$.
Furthermore, the received signals
\begin{equation}
  y_n(t)=\alpha_{mn}s_{mr}(t)+n(t)	
\end{equation}
are corrupted by a complex additive white Gaussian noise $n(t)$ with
variance $1/2$ per dimension.

 At the receiver, the detection is done on each of the $L_r$
received signals separately. Therefore, in general, each code block
$\mathbf S(t)$ has to be detected by block. E.g. for a 2x2 block,
estimating the symbols $\hat d_j$ implies \lucout{an effort}
\lucin{computational complexity} proportional to $M^2$. Now, this
\lucout{effort}\lucin{complexity} can be reduced to $2M$ by
\lucout{introducing some properties}\lucin{introducing an
  orthogonality property as well as simplifying assumptions} on the
code.

Criteria for such ST\lucin{B}C are given in Section \ref{sec:Cond}. In
Section \ref{sec:OSTC}, the criteria are used to construct
OST\lucin{B}C for CPM. In Section \ref{sec:simul1} we test the designed
code and compare it with the STC from Wang and Xia \cite{Wang05}.
Finally, some conclusions are drawn in Section \ref{sec:concl1}.

\subsection{Design Criteria}
\label{sec:Cond}

The purpose of the design is to achieve full diversity and a fast
maximum likelihood decoding while maintaining the continuity of the
signal phases.  \lucin{This section shows how the need to perform fast
  ML decoding leads to the $L^2$ orthogonality condition as well as to
  simplifying assumptions, which can be combined with the continuity
  conditions.} For convenience we only consider one Rx antenna and drop the 
  index $n$ in $\alpha_{mn}$.

\lucin{\subsubsection{Fast Maximum Likelihood Decoding}} 

Commonly, due to the trellis structure of CPM, the Viterbi algorithm is used 
to perform the ML demodulation. On block $l$ each state in the trellis has $M^2$ incoming
branches and $M^2$ outgoing branches with a distance

\begin{equation}
  D_l= \int\limits_{2lT}^{(2l+1)T} \Big|
      y(t)-\sum\limits_{m=1}^2 \alpha_{m}s_{m1}(t) \Big| ^2\D
    t+
    \int\limits_{(2l+1)T}^{(2l+2)T} \Big|
      y(t)-\sum\limits_{m=1}^2 \alpha_{m}s_{m2}(t) \Big|^2\D t.
  \label{eq:MLext}
\end{equation}
The number of branches results from the blockwise decoding and the
 correlation between the sent symbols $s_{1r}(t)$ and $s_{2r}(t)$. 
 A way to reduce the number of branches  is to
structurally decorrelate the signals sent by the two transmitting
antennas, i.e. to put to zero the inter-antenna correlation
\begin{equation}
  \alpha_{2}\alpha_{1}^*\!\!\!\!\!\!\int\limits_{2lT}^{(2l+1)T}
  \!\!\!\!\!\! s_{21}(t)s_{11}^*(t)\D t+\alpha_{1}\alpha_{2}^*\!
  \!\!\!\!\!\!\! \int\limits_{2lT}^{(2l+1)T} \!\!\!\!\!\!\!\!
  s_{11}(t)s_{21}^*(t)\D t + \alpha_{2}\alpha_{1}^*
  \!\!\!\!\!\!\!\int\limits_{(2l+1)T}^{(2l+2)T}\! \!\!\!\!\!\!
  s_{22}(t)s_{12}^*(t)\D t+\alpha_{1}\alpha_{2}^*\!
  \!\!\!\!\!\!\int\limits_{(2l+1)T}^{(2l+2)T}\! \!\!\!\!\!\!
  s_{12}(t)s_{22}^*(t)\D t = 0.
  \label{eq:cross0}
\end{equation}

Pointwise orthogonality as defined in \cite{Wang05} is therefore a
sufficient condition but not necessary. A less restrictive $L^2$
orthogonality is also sufficient. From Eq. (\ref{eq:cross0}), the distance 
given in Eq.  (\ref{eq:MLext}) can then be simplified to

\begin{equation}
  D_l=\int\limits_{2lT}^{(2l+1)T}\!\!\!\!\!\!\!
    f_{11}(t)+f_{21}(t)-|y(t)|^2 \D t+\int\limits_{(2l+1)T}^{(2l+2)T}\!\!\!\!\!\!\!
    f_{12}(t)+f_{22}(t) -|y(t)|^2 \D t 
\end{equation}
with $f_{mr}(t)=|y(t)-\alpha_m s_{mr}(t)|^2$. When each $s_{mr}(t)$
depends only on $d_{2l+1}$ or $d_{2l+2}$ the branches can be split and 
calculated separately  for  $d_{2l+1}$ and $d_{2l+2}$. The
complexity of the ML decision is reduced to $2M$. The complexity for
detecting two symbols is thus reduced from $pM^{\gamma+1}$ to
$pM^{\gamma}$. The STC introduced by Wang and Xia \cite{Wang05}
didn't take full advantage of the orthogonal design since $s_{mr}(t)$
was depending on both $d_{2l+1}$ and $d_{2l+2}$. The gain they
obtained in \cite{Wang05} was then relying on other properties of CPM,
e.g. some restrictions put on $q(t)$ and $p$. These
restrictions may also be applied to our design code, which would lead
to additional complexity reduction.

\subsubsection{Orthogonality Condition}

In this section we show how $L^2$ orthogonality for CPM, i.e.
$\|\mathbf S(t)\|^2_{L_2}=\int_{2lT}^{(2l+2)T}\mathbf S(t)\mathbf S^H(t)\D t=2\mathbf I$, can
be obtained. As such, the correlation between the two transmitting
antennas per coding block is canceled if

\begin{equation}
  \int\limits_{2lT}^{(2l+2)T}s_{1r}(t)s_{2r}^*(t)\D
  t=\int\limits_{2lT}^{(2l+1)T}s_{11}(t)s_{21}^*(t)\D t 
  +  \int\limits_{(2l+1)T}^{(2l+2)T}s_{12}(t)s_{22}^*(t)\D t=0\text{.}
\end{equation}
 Replacing $s_{mr}(t)$ by the corresponding CPM symbols from Eq.
(\ref{eq:phase}), we get\small
\begin{multline}
     \!\int\limits_{2lT} ^{(2l+1)T}\!\!\!\!\!\!\!  \exp\!\Big\{j2\pi\big[\theta_1(2l+1)+   h\!\!\!\!\!\!\!\sum\limits_{i=2l+2-\gamma}^{2l+1}\!\!\!\!\!\!\! d_{1,1}^{(i)} q(t-(i-1)T)+c_{1,1}(t)-
       \theta_2(2l+1)-h\!\!\!\!\!\!\sum\limits_{i=2l+3-\gamma}^{2l+2}\!\!\!\!\!\! d_{2,1}^{(i)}q(t-(i-1)T)-c_{2,1}(t)\big]\Big\}\D t+ \\
            \!\int\limits_{(2l+1)T} ^{(2l+2)T} \!\!\!\!\!\!\!\exp\!\Big\{j2\pi\big[\theta_1(2l+2)+h\!\!\!\!\!\!\!\sum\limits_{i=2l+3-\gamma}^{2l+2}\!\!\!\!\!\!\!d_{1,2}^{(i)}q(t-(i-1)T)\!+\!c_{1,2}(t)-
     \theta_2(2l+2)-h\!\!\!\!\!\!\sum\limits_{i=2l+2-\gamma}^{2l+1}  \!\!\!\!\!\!d_{2,2}^{(i+1)}q(t-iT) -c_{2,2}(t) \big]\D t\Big\} = 0.
\end{multline}\normalsize
The phase memory $\theta_m(2l+r)$ is independent of time and has not
to be considered for integration. Using Eq. (\ref{eq:theta1}) to
replace phase memory $\theta_m(2l+2)$ of the second time slot, we
obtain\small
\begin{multline}
  \int\limits_{2lT} ^{(2l+1)T} \!\!\!\!\!\!
  \exp\Big\{j2\pi\big[h\!\!\!\!\!\!\sum\limits_{i=2l+2-\gamma} ^{2l+1}
  \!\!\!\!\!\!  d_{1,1}^{(i)}q(t-(i-1)T) +c_{1,1}(t)-
  h\!\!\!\!\!\!\sum\limits_{i=2l+2-\gamma}^{2l+1}
  \!\!\!\!\!\!d_{2,1}^{(i)}q(t-(i-1)T)-c_{2,1}(t)\big]\Big\}\D t+
  \exp\Big\{ j2\pi\big[\xi_1(2l+1)-\xi_2(2l+1)\big]\Big\}\cdot\\
  \int\limits_{2lT}^{(2l+1)T}\!\!\!\!\!\!\! \exp
  \Big\{j2\pi\big[h\!\!\!\!\!\!\sum\limits_{i=2l+2-\gamma}^{2l+1}\!\!\!\!\!\!d_{1,2}^{(i+1)}
  q(t-(i-1)T)+c_{1,2}(t+T)-
  h\!\!\!\!\!\!\sum\limits_{i=2l+2-\gamma}^{2l+1}
  \!\!\!\!\!\!d_{2,2}^{(i+1)}q(t-(i-1)T) -c_{2,2}(t+T) \big]\Big\}\D t
  = 0.
  \label{eq:orthLong}
\end{multline}\normalsize

\lucin{\subsubsection{Simplifying assumptions}} To simplify this
expression, we \lucin{factor} \lucout{split} Eq. (\ref{eq:orthLong}) into a time
independent and a time dependent part. For merging the two integrals
to one time dependent part, we have to map $d_{m2}^{(i)}$ to
$d_{m1}^{(i)}$ and $c_{mr}(t)$ to a different
$c_{m\lucin{'}r\lucin{'}}(t)$. Consequently, for the data symbols
$d_{mr}^{(i)}$ there exist three possible ways of mapping:
\begin{itemize}
\item {\em crosswise mapping} with $d_{1,1}^{(i)}=d_{2,2}^{(i)}$ and
  $d_{1,2}^{(i)}=d_{2,1}^{(i)}$;
\item {\em repetitive mapping} with   $d_{1,1}^{(i)}=d_{1,2}^{(i)}$ and
  $d_{2,1}^{(i)}=d_{2,2}^{(i)}$;
\item {\em parallel mapping} with $d_{1,1}^{(i)}=d_{2,1}^{(i)}$ and
  $d_{1,2}^{(i)}=d_{2,2}^{(i)}$ .
\end{itemize}
The same approach can be applied to $c_{mr}(t)$:
\begin{itemize}
\item {\em crosswise mapping} with $c_{11}(t)=-c_{22}(t-T)$ and
  $c_{12}(t)=-c_{21}(t-T)$;
\item {\em repetitive mapping} with $c_{11}(t)=c_{12}(t-T)$ and 
$c_{21}(t) =c_{22}(t-T)$;
\item {\em parallel mapping} with $c_{11}(t)=c_{21}(t)$ and
  $c_{12}(t)=c_{22}(t)$.
\end{itemize}

 For each combination of mappings, Eq. (\ref{eq:orthLong}) is now
\lucout{ in two products}\lucin{the product of two factors}, one
containing the integral and \lucin{the other} a time independent part.
To fulfill Eq. (\ref{eq:orthLong}) it is sufficient if one factor is
zero, namely $ 1+\ee^{j2\pi\left[\xi_1(2l+1)-\xi_2(2l+1)\right]} =0$,
i.e. if
\begin{equation}
   k+\frac{1}{2}=\xi_1(2l+1)-\xi_2(2l+1)
  \label{eq:orthCond}
\end{equation}
with $k\in \mathbb N$. We thus get a very simple condition which only
depends on $\xi_m(2l+1)$.

\subsubsection{Continuity of Phase}

In this section we determine the functions $\xi_m(2l+1)$ to ensure the
phase continuity.

Precisely,  the phase of the CPM symbols has to be equal at all
intersections of symbols. For an arbitrary block $l$, it means that
$\phi_{m1}((2l+1)T) =\phi_{m2}((2l+1)T)$. Using Eq. (\ref{eq:phase}),
it results in
\begin{equation}
  \xi_m(2l+1) = h \!\!\!\!\!\!\!\!\!\sum\limits_{i=2l+2-\gamma}^{2l+1}
  \!\!\!\!\!\!\!\! d_{m,1}^{(i)}q((2l+2-i)T)+c_{m,1}((2l+1)T) 
-h \!\!\!\!\!\!\!\sum\limits_{i=2l+3-\gamma}^{2l+2}
  \!\!\!\!\!\!\! d_{m2}^{(i)}q((2l+2-i)T)-c_{m,2}((2l+1)T)\text{.}
  \label{eq:xi1org}
\end{equation}
 For the second intersection at $(2l+2)T$, since
$\phi_{m2}((2l+2)T)=\phi_{m1}((2l+2)T)$, we get
\begin{equation}
  \xi_m(2l+2)=h \!\!\!\!\!\!\sum\limits_{i=2l+3-\gamma}^{2l+2}
  \!\!\!\!\!\!\! d_{m2}^{(i)}q((2l+3-i)T)+c_{m2}((2l+2)T)-
h  \!\!\!\!\!\!\!\!\!\!\sum\limits_{i=2(l+1)+2-\gamma}^{2(l+1)+1}
  \!\!\!\!\!\!\!\!\!\!\!
  d_{m,1}^{(i)}q((2l+3-i)T)-c_{m,1}((2l+2)T)\text{.}
  \label{eq:xi2org}
\end{equation}

 Now, by choosing one of the mappings detailed in Section
\ref{sec:OSTC}, these equations can be greatly simplified. Hence, we
have all the tools to construct our code.

\subsection{Orthogonal Space Time Codes}
\label{sec:OSTC}

In this section we will have a closer look at two codes constructed
under the afore-mentioned conditions.

\subsubsection{Existing Code}

As a first example, we will give an alternative construction of the
code given by Wang and Xia in \cite{Wang05}.  \lucin{Indeed, the
  pointwise orthogonality condition used by Wang and Xia is a special
  case of the $L^2$ orthogonality condition, hence, their ST-code can
  be obtained within our framework.} \lucout{ Namely, as expected, this
  ST-code can be obtained within our framework. It is easily verified
  to be orthogonal in $L^2$ sense but this is not surprising since
  Wang and Xia constructed their code from pointwise orthogonality
  (Alamouti scheme) and this is a special case of $L^2$
  orthogonality.}

For the first antenna Wang and Xia use a conventional CPM with
$d_{1r}^{(i)}=d_i$ for $i=2l+r+1-\gamma,2l+r+2-\gamma,\ldots ,2l+r$
and $c_{1r}(t)=0 $. The symbols of the second antenna are mapped
{ \em crosswise} to the first $d_{21}^{(i)}=-d_{i+1}$ for
$i=2l+2-\gamma,2l+3-\gamma,\ldots ,2l+1$ and $d_{22}^{(i-1)}=-d_{i-1}$
for $i=2l+3-\gamma,2l+4-\gamma,\ldots ,2l+2$. Using this cross mapping
makes it difficult to compute $\xi_m(2l+1)$ since the CPM typical
order of the data symbols is mixed. Wang and Xia circumvent this by
introducing another correction factor for the second antenna
\begin{equation}
  c_{2r}(t)\!=\!\!\sum\limits_{i=0}^{\gamma-1}(h(
  {d_{2l+1-i}+d_{2l+2-i}})+1 ) q_0(t-(2l+r-1-i)T).
\end{equation}
By first computing $\xi_m(2l+1)$ with Eq.  (\ref{eq:xi1}) and then Eq.
(\ref{eq:orthCond}), we get the $L^2$ orthogonality of the
Wang-Xia-STC.

\subsubsection{Parallel Code}

To get a simpler correction factor as in \cite{Wang05}, we
designed a new code based on the {\em parallel} structure which permits
to maintain the conventional CPM mapping for both antennas. Hence we
choose the following mapping: $d_{m1}^{(i)}=d_{m2}^{(i-1)}=d_i$ for
$i=2l+r+1-\gamma,2l+r+2-\gamma,\ldots ,2l+r$. Then, Eq.
(\ref{eq:xi1org}) and (\ref{eq:xi2org}) can be simplified into
\begin{equation}
  \xi_m(2l+1)=\frac{h}{2}d_{2l+2-\gamma}+c_{m1}((2l+1)T) -c_{m2}((2l+1)T) \label{eq:xi1}
  \xi_m(2l+2)=\frac{h}{2}d_{2l+3-\gamma}+c_{m2}((2l+2)T) -c_{m1}((2l+2)T)\text{.}
\end{equation}
 With this simplified functions, the orthogonality condition
only depends on the start and end values of $c_{mr}(t)$, i.e.
\begin{equation}
  k+\frac{1}{2}=c_{11}((2l+1)T)-c_{12}((2l+1)T)
 -c_{21}((2l+1)T)+c_{22}((2l+1)T).
  \label{eq:condC}
\end{equation}
To merge the two integrals in Eq. (\ref{eq:orthLong}), the mapping of $d_{mr}^{(i)}$ is necessary but also an equality
between different $c_{mr}(t)$. From the three possible mappings, we
choose the {\em repeat mapping} because of the possibility to set
$c_{mr}(t)$ to zero for one antenna. Hence we are able to send a
conventional CPM signal on one antenna and a modified one on the
second.  Using Eq.  (\ref{eq:condC}) and the equalities for the
mapping, we can formulate the following condition
\begin{equation}
  k+\frac{1}{2}=c_{12}(2lT)-c_{12}((2l+1)T)-c_{22}(2lT)+c_{22}((2l+1)T).
  \label{eq:cR}
\end{equation}


 With $c_{11}(t)=c_{12}(t)=0$, we can take for
$c_{21}(t)=c_{22}(t)$ any continuous function which is zero at $t=0$
and $1/2$ at $t=T$.  Another possibility is to choose the correction
factor of the second antenna with a structure similar to CPM
modulation, i.e.
\begin{equation}
  c_{2r}(t)=\sum\limits_{i=2l+1-\gamma}^{2l+1}q(t-(i-1)T)
\end{equation}
for ${(2l+r-1)T\leq t\leq(2l+r)T}$.  With this approach, the
correction factors can be included in a classical CPM modulation with
constant offset of $1/h$. This offset may also be expressed as a
modified alphabet for the second antenna
\begin{equation}
  \Omega_{d_2}=\{ -M+1+\frac{1}{h},-M+3+\frac{1}{h},\ldots,M-3+\frac{1}{h},M-1+\frac{1}{h}\}.
  \label{eq:omega2}
\end{equation}

Consequently, this $L^2$-orthogonal design may be seen as two
conventional CPM designs with different alphabet sets $\Omega_d$ and
$\Omega_{d2}$ for each antenna. However, in this method, the constant
offset to the phase may cause a shift in frequency. But as shown by
our simulations in the next section, this shift is quite moderate.

\subsection{Simulations}
\label{sec:simul1}

\begin{figure}
\begin{tabular}{cc}
  \includegraphics[width=3.1in]{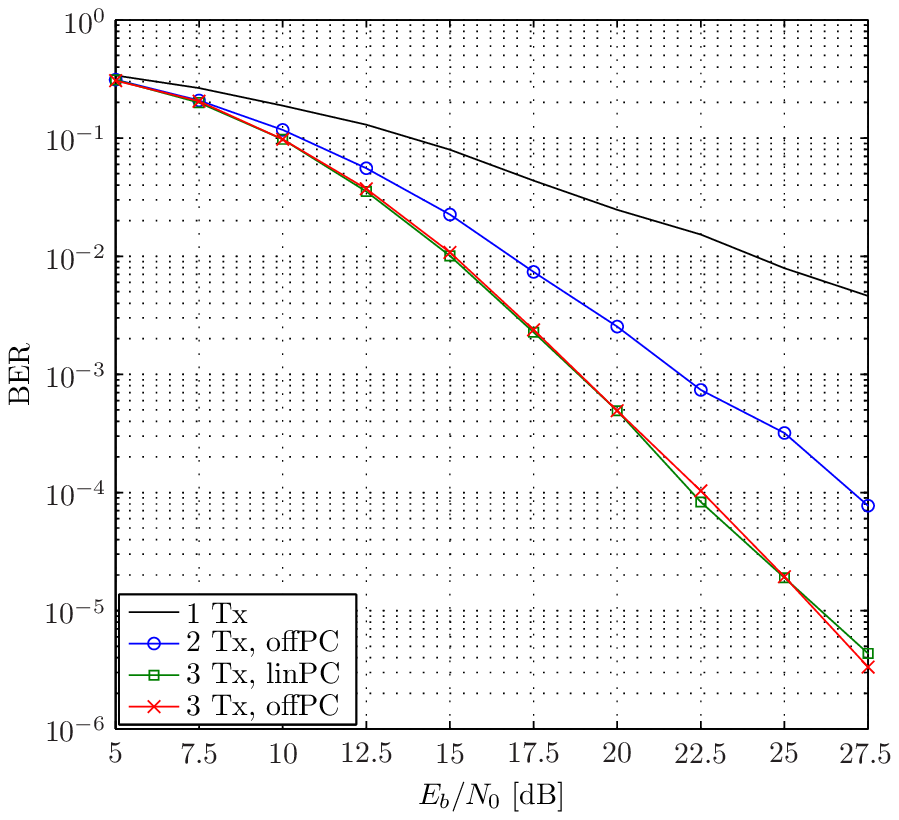}&\includegraphics[width=3.2in]{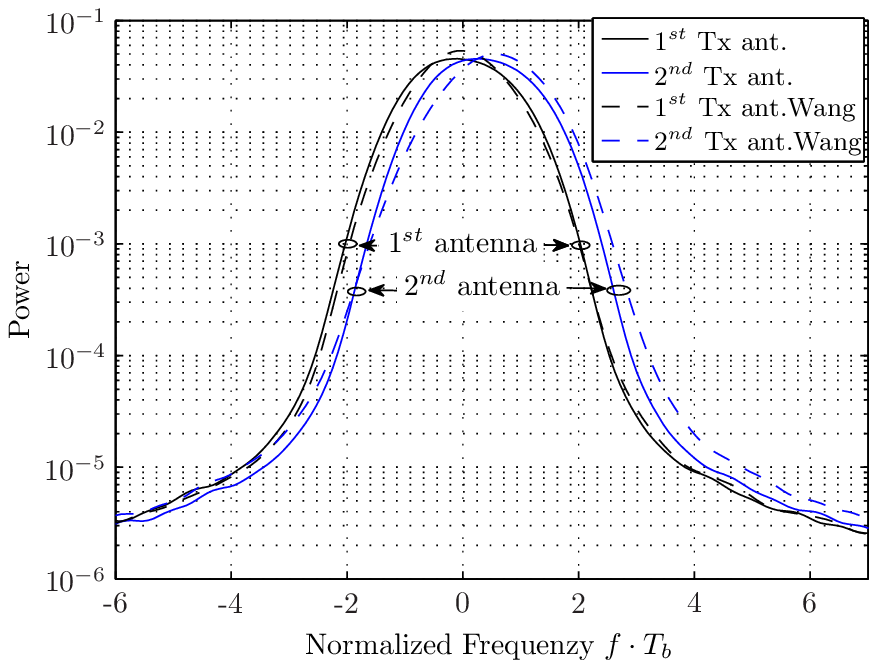}
\end{tabular}
  \caption{\small Left: Simulated BER for different numbers of Tx and Rx antennas of
    the proposed STC and of the Wang-Xia-STC; Right: Simulated psd for each Tx antenna of the proposed STC
    (continuous line) and the Wang-Xia-STC (dashed line) }
  \label{fig:BER1}\label{fig:Spec1}
\end{figure}

In this section we verify the proposed algorithm by simulations.
Therefore a STC-2REC-CPM-sender with two transmitting antennas has
been implemented in MATLAB. For the signal of the first antenna we use
conventional Gray-coded CPM with a modulation index $h=1/2$, the
length of the phase response function $\gamma=2$ and an alphabet size
of $M=8$. The signal of the second antenna is modulated by a CPM with
the same parameters but a different alphabet $\Omega_{d_2}$,
corresponding to Eq. (\ref{eq:omega2}).

The channel used is a frequency flat Rayleigh fading model with
additive white Gaussian noise. The fading coefficients $\alpha_{mn}$
are constant for the duration of a code block
(\lucout{fast}\lucin{block} fading) and known at receiver (coherent
detection). The received signal $y_n(t)$ is demodulated by two
filterbanks with $pM^2$ filters, which are used to calculate the
correlation between the \lucout{truly} received and \lucout{possible
  received}\lucin{candidate} signals. Due to the orthogonality of the
antennas each filterbank is independently applied to the corresponding
time slot $k$ of the block code. The correlation is used as metric for
the Viterbi algorithm (VA) which has $pM$ states and $M$ paths
leaving each state.  \lucout{For}\lucin{In our} simulation\lucin{,}
the VA is truncated to a path memory of 10 code blocks, which means
that we get a decoding delay of $2\cdot 10T$.

\lucout{Figure \ref{fig:BER1} shows the results of our simulations. The
  continuous line without marker is the result of a system without
  diversity (1 transmitting (Tx) and 1 receiving (Rx) antenna) and is
  used as reference. The curve marked with circles results from the
  coding introduced by Wang and Xia \cite{Wang05} using 2 Tx and 1 Rx
  antenna.  The plots marked with buttons and triangles we obtained by
  the coding introduced in this part with 2 Tx and 1 or 2 Rx
  antennas, respectively.

  As expected, the use of multiple independent channels yields an
  improved BER behavior. The reference system with one channel (1 Tx
  and 1 Rx antenna) has a slope of slightly more than 10dB per decade
  for high EbNo. For 2 Tx antennas and 1 Rx antenna we obtain a slope
  of around 5 dB per decade for both the STC of Wang and Xia and our
  approach. A doubling of the slope while doubling transmitter
  antennas at the same time, corresponds to full diversity. The small
  difference between the proposed STC and the one of Wang and Xia may
  be explained by our simpler design. Finally, by doubling the number
  of receiving antennas we reach a steepness of a bit less than 3 dB
  per decade, which approximately corresponds to full diversity. Due
  to these simulation results we can reasonably assume that the
  proposed code achieves full diversity.  }

\lucin{From the simulation results given in Figure \ref{fig:BER1}, we
  can reasonably assume that the proposed code achieves full
  diversity.  Indeed, the curves for the 2x1 and 2x2 systems
  respectively show a slope of 2 and 4.  Moreover, the curve of the
  2x1 systems follows the same slope as the ST code proposed by Wang
  and Xia \cite{Wang05}, which was proved to have full diversity.
  Note also that the new code provides a slightly better performance.}

A main reason of using CPM for STC is the spectral efficiency. Figure
\ref{fig:Spec1} show the simulated power spectral density (psd) for
both Tx antennas of the proposed ST code (continuous line) and the ST
code proposed by Wang and Xia \cite{Wang05}. The first antenna of our
approach uses a conventional CPM signal and hence shows an equal psd.
The spectrum of the second antenna is shifted due to adding an offset
$c_{mr}(t)$ with a non zero mean. Minimizing the difference between
the two spectra by shifting one, result in a phase difference of
$0.375$ measured in normalized frequency $f\cdot T_d$, where
$T_d=T/\log_2(M)$ is the bit symbol length. The first antenna of the
Wang-Xia-algorithm has almost the same psd while the spectrum of the
second antenna is shifted by approximately $0.56 f\cdot T_d$. This
means that the OSTC by Wang and Xia requires a slightly larger
bandwidth than our OSTC.

\subsection{Conclusion to part one}
In applications where the power efficiency is crucial, combination of
Continuous Phase Modulation and Space Time Coding has the potential to
provide high spectral efficiency, thanks to spatial diversity. To
address this power efficiency, ST code design for CPM has to
ensure both low complexity decoding and full diversity.  To fulfill
these requirements, we have proposed a new $L^2$ orthogonality
condition.  We have shown that this condition is sufficient to achieve
low complexity ML decoding and leads, with the help of simplifying
assumption to a simple code.  Moreover, simulations indicate that the
code most probably achieves full diversity. In the next part of this
report, we will
concentrate on the design of other codes based on $L^2$ orthogonality
and will show how to  design full diversity, 
full rate $L^2$ orthogonal codes for 3 antennas.
\label{sec:concl1}

\section{Part two: Extension to more antennas}
To combine the power efficiency of Continuous Phase Modulation (CPM)
with enhanced performance in fading environments, some authors have
suggested to use CPM in combination with Space-Time Codes (STC).
In part one, we have proposed a CPM ST-coding scheme bases on
$L^2$-orthogonality for two transmitting antennas. In this part we
extend this approach to the three antenna case. We analytically derive
a family of coding schemes which we call Parallel Code (PC).  This
code family has full rate and we expect that the proposed coding
scheme achieves full diversity. This is confirmed by accompanying
simulations. We detail an example for the proposed STC which can be
interpreted as a conventional CPM scheme with different alphabet sets
for the different transmit antennas which results in simplified
implementation.  Thanks to $L^2$-orthogonality, the decoding
complexity, usually exponentially proportional to the number of
transmitting antennas, is reduced to linear complexity.



\subsection{Introduction}
\label{sec:intro}
To overcome the reduction of channel capacity caused by fading,
Telatar \cite{Tela99}, Foschini and Gans \cite{Fosc98} described in
the late 90s the potential gain of switching to multiple input
multiple output (MIMO) systems.  These results triggered many advances
mostly concentrated on the coding aspects for transmitting antennas,
e.g.  Alamouti \cite{Alam98} and Tarokh et al.  \cite{Taro99a} for
Space-Time Block Codes (STBC) and also Tarokh et al. \cite{Taro98} for
Space-Time Trellis Codes.

\begin{figure}
  \centering \includegraphics[width=0.5\textwidth]{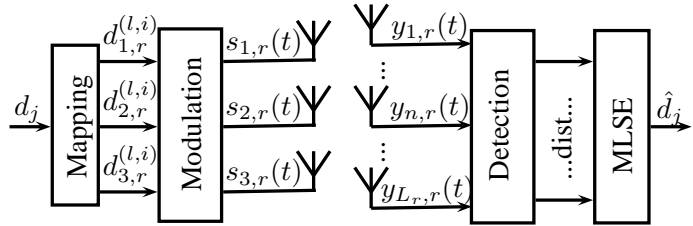}
  \caption{Structure of a MIMO Tx/Rx system }
  \label{fig:block}
\end{figure}

Zhang and Fitz \cite{Zhan00,Zhan03} were the first to apply the idea
of STC to CPM by constructing trellis codes.  In \cite{Silv06},
Silvester et al.  derived a diagonal block space-time code which
enables non-coherent detection. A condition for optimal coding gain
while sustaining full diversity was also recently derived by Zaji\'c
and St\"uber \cite{Zaji07}.

Inspired by orthogonal design codes, Wang and Xia introduced in
\cite{Wang04} the first orthogonal STC for two transmitting antennas
and full response CPM and later in \cite{Wang05} for partial response.
Their approach was extended in \cite{Wang03} to construct a
pseudo-orthogonal ST-coded CPM for four antennas. To avoid the
structural limitation of orthogonal design, we proposed in
\cite{Hess08} a STC CPM scheme based on $L^2$ orthogonality for two
antennas. Sufficient conditions for $L^2$ orthogonality were
described, $L^2$ orthogonal codes were introduced and the simulation
results displayed good performance and full rate. Here, motivated by
this results, we extend our previous work and generalize these
conditions for three transmitting antenna.

The main result of the three transmit antenna case, is that it can,
unlike the codes based on orthogonal design, achieve full diversity
with a full rate code:
 
\begin{enumerate}
\item {\bf the full rate property} is one of the main advantage of
  using the $L^2$ norm criterion, instead of merely extending the
  classical Tarokh \cite{Taro99a} orthogonal design to the CPM case.
  Indeed, in the classical orthogonal design approach, which is based
  on optimal decoding for linear modulations, the criterion is
  expressed as the orthogonality between matrices of elements, each of
  these elements being a definite integral (usually the output of a
  matched filter).  On the contrary, in the $L^2$ design approach used
  for non-linear modulations, the product in Eq.  (\ref{eq:condLong})
  is a definite integral itself, the integrand being the product of
  two signals. This allows more degrees of freedom and enables the
  full rate property.

\item {\bf the full diversity property} can be proved in a similar way
  to the classical case, with the help of the extensions proposed by
  Zaji\'c and St\"uber \cite{Zaji07}.
\end{enumerate}

Furthermore, it should be pointed out that the proposed coding scheme
does not limit any parameter of the CPM. It is applicable to full and
partial response CPM as well as to all modulation indexes.

We first give the system model for a multiple input multiple output
(MIMO) system with $L_t$ transmitting (Tx) antennas and $L_r$
receiving (Rx) antennas (Fig. \ref{fig:block}). Later on, we will use
this general model to derive a L2-OSTC for CPM for $L_t=3$.  The
emitted signals $\mathbf s(t)$ are mixed by a channel matrix $\mathbf
A$ of dimension $L_r \times L_t$. The elements of $\mathbf A$,
$\alpha_{n,m}$, are Rayleigh distributed random variables and
characterize the fading between the $n^{th}$ Rx and the $m^{th}$ Tx
antenna. The Tx signal is disturbed by complex additive white Gaussian
noise (AWGN) with variance of $1/2$ per dimension which is represented
by a $L_r \times L_t$ matrix $ \mathbf n(t)$. The received signal
\begin{equation}
  \mathbf y(t)=\mathbf A\mathbf s(t)+\mathbf n(t).
\end{equation}
has the elements $y_{n,r}$ and the dimension $L_r \times L_t$. We
group the transmitted CPM signals into blocks
\begin{equation}
  \mathbf s(t)=%
  \begin{bmatrix}
    s_{1,1}(t)&\ldots&s_{1,L_t}(t)\\
    \vdots &s_{m,r}(t)&\vdots\\
    s_{L_t,1}(t)&\ldots&s_{L_t, L_t}(t)\\
  \end{bmatrix}
  \label{eq:smatrix}
\end{equation}
similar to a ST block code with the difference that now the elements
are functions of time and not constant anymore. The elements of Eq.
(\ref{eq:smatrix}) are given by
\begin{equation}
  s_{m,r}(t)=\sqrt{\frac{E_s}{L_tT}}\exp\left( j2\pi\phi_{m,r}(t)\right)
  \label{eq:defs}
\end{equation}
for $(L_tl+r-1)T\leq t\leq (L_tl+r)T$ and $m,r=1,2,\ldots,L_t$. Here
$m$ represents the transmitting antenna and $r$ the relative time slot
in the block. The symbol energy $E_s$ is normalized to the number of
Tx antennas $L_t$ and the symbol length $T$. The continuous phase
\begin{equation}
  \phi_{m,r}(t)=\theta_m(L_tl+r)+h\sum_{i=1}^{\gamma}d_{m,r}^{(l,i)}q(t-i'T)+c_{m,r}(t)
  \label{eq:defphi}
\end{equation}
is defined similarly to \cite{Ande86} with an additional correction
factor $c_{m,r}(t)$ detailed in Section \ref{sec:CorrF}.  Furthermore,
$l$ is indexing the whole code block, $i$ the overlapping symbols for
partial response and $i'=L_tl+r-i$. With this extensive description of
the symbol $d_{m,r}^{(l,i)}$, we are able to define all possible
mapping schemes (cp. Section \ref{sec:mapping}).  The modulation index
$h=2m_0/p$ is the quotient of two relative prime integers $m_0$ and
$p$ and the phase smoothing function $q(t)$ has to be continuous for
$0\leq t\leq\gamma T$, $0$ for $t\leq0$ and $1/2\leq\gamma T$. The
memory length $\gamma$ gives the number of overlapping symbols.

To maintain continuity of phase, we define the phase memory
\begin{equation}
  \theta_m(L_tl+r+1)=\theta_m(L_tl+r)+\xi_m(L_tl+r)
  \label{eq:theta}
\end{equation}
in a general way. The function $\xi(L_tl+r)$ will be fully defined in
Section \ref{sec:CorrF} from the contribution of $c_{mr}(t)$. For a
conventional CPM system, we have $c_{mr}(t)=0$ and
$\xi(2l+1)=\frac{h}{2}d_{2l+1-\gamma}$.


In Section \ref{sec:PCS}, we derive the $L^2$ conditions for a CPM
with three transmitting antennas and introduce adequate mappings and a
family of correction factors. In Section \ref{sec:propert}, we detail
some properties of the code. In Section \ref{sec:simul}, we benchmark
the code by running some simulations and finally, in Section
\ref{sec:concl}, some conclusions are drawn.

\subsection{Parallel Codes (PC) for 3 antennas}
\label{sec:PCS}
\subsubsection{$L^2$ Orthogonality}
\label{sec:l2ortho}

In this section we describe how to enforce $L^2$ orthogonality on CPM
systems with three transmitting antennas. Similarly to \cite{Hess08},
we impose $L^2$ orthogonality by
\begin{equation}
  \int\limits_{3lT}^{(3l+3)T}\mathbf s(t)\mathbf s^H(t)\D t=E_S\mathbf I
\end{equation}
where $\mathbf I$ is the $3\times3$ identity matrix. Hence the
correlation between two different Tx antennas $s_{m,r}(t)$ and
$s_{m',r}(t)$ is canceled over a complete STC block if
\begin{equation}
  \int\limits_{3lT}^{(3l+3)T}s_{m,r}(t)s_{m',r}^*(t)\D t = 0
\end{equation}
with $m\neq m'$. Now, by using Eq. (\ref{eq:defs}) and
(\ref{eq:defphi}) we get
\begin{equation}
  0=\sum_{r=1}^{3}\int\limits_{(3l+r-1)T}^{(3l+r)T}\exp\Big(j2\pi\cdot
  \big[ \theta_m(3l+r)+h\sum_{i=1}^\gamma
  d_{m,r}^{(l,i)}q(t-i'T)+c_{m,r}(t)-
  (\theta_{m'}(3l+r)-h\sum_{i=1}^\gamma
  d_{m',r}^{(l,i)}q(t-i'T)-c_{m',r}(t))\big]\Big)\D t.
  \label{eq:condLong}
\end{equation}
The phase memory $\theta_m(3l+r)$ is time independent and therewith
can be moved to a constant factor in front of the integrals. Similarly
to \cite{Hess08}, we introduce {\em parallel mapping}
$(d_{m,r}^{(l,i)}=d_{m',r}^{(l,i)})$ for the data symbols and {\em
  repetitive mapping} $(c_{m,r}(t)=c_{m,r'}(t))$ for the correction
factors. The integral on three time slots can then be merged into one
time dependent factor.  Furthermore, we obtain a second, time
independent factor from the phase memory. Now, by using Eq.
(\ref{eq:theta}) one can see that the condition from Eq.
(\ref{eq:condLong}) is fulfilled if
\begin{equation}
  \label{eq:cond}
  0 = 1 + \exp(ja_1) +  \exp(ja_1) \exp(ja_2)
\end{equation}
where $a_r=2\pi\left[ \xi_m(3l+r)-\xi_{m'}(3l+r)\right]$ and we get
$-\exp(-ja_1)=1+\exp(ja_2)$. By splitting this equation into imaginary
and real parts, we have the following two conditions:
\begin{align}
  -1=&\cos(-a_1)+\cos(a_2)\\
  0=&\sin(-a_1)+\sin(a_2).
\end{align}
This system has, modulo $2\pi$, two pairs of solutions
\begin{equation}
  (a_1,a_2)\in\{(2\pi/3,2\pi/3),(4\pi/3,4\pi/3)\}.
\end{equation}

Hence $L^2$ orthogonality is achieved if $
\xi_m(3l+r)-\xi_{m'}(3l+r)=1/3$ or $ \xi_m(3l+r)-\xi_{m'}(3l+r)=2/3$
for $r=1,2$ and for all combinations of $m$ and $m'$ with $m\neq m'$.
In order to determine $\xi_m(3l+r)$, we detail in the following
section the exact mapping and the correction factor.



\label{sec:pc}
\subsubsection{Mapping}
\label{sec:mapping}

\begin{figure}
  \centering \includegraphics[width=0.4\textwidth]{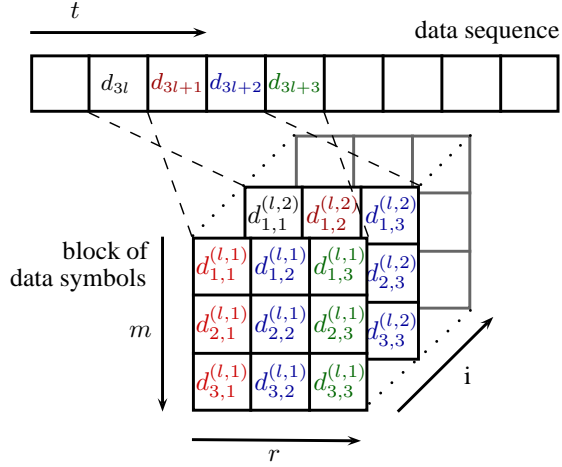}
  \caption{Mapping of the data sequence to the data symbols }
  \label{fig:mapping}
\end{figure}

In this section we describe the mapping of the data sequence $d_j$ to
the data symbols $d_{m,r}^{(l,1)}$ of the block code (Fig.
\ref{fig:mapping}). To obtain full rate each code block have to
include three new symbols from the data sequence. In general, the
mapping of the three new symbols has no restrictions. However, to fix
a mapping two criteria are considered:
\begin{itemize}
\item mapping to simplify Eq. (\ref{eq:condLong})
\item low complexity of function $\xi_m(3l+r)$.
\end{itemize}
The first criteria is already determined by using {\em parallel
  mapping} $(d_{m,r}^{(l,i)}=d_{m',r}^{(l,i)})$. Therewith the mapping
for the $m$-dimension (Fig.  \ref{fig:mapping}) is fixed. For the
remaining two dimensions we choose a mapping similar to conventional
CPM. The subsequent data symbols in $r$-direction are mapped to
subsequent symbols from the data sequence. Also similar to
conventional CPM we shift this mapping by $-i$ and obtain
\begin{equation}
  d_{m,r}^{(l,i)}=d_{3l+r-i+1}.
\end{equation}
Eq. (\ref{eq:xiLong}) and (\ref{eq:xi}) show the simplification of the
function $\xi_m(3l+r)$.

\subsubsection{Correction Factor}
\label{sec:CorrF}

The choice of the phase memory and therewith of the function
$\xi_m(3l+r)$ ensures the continuity of phase. If
$\phi_{m,r}((L_tl+r)T)=\phi_{m,r+1}((L_tl+r)T)$, we always obtain the
desired continuity. Hence,
\begin{equation}
  \xi_m(L_tl+r)=h\sum\limits_{i=1}^{\gamma}d_{m,r}^{(l,i)}q(iT)+c_{m,r}((3l+r)T)-
  h\sum\limits_{i=1}^{\gamma}d_{m,r+1}^{(l,i)}q(iT)-c_{m,r+1}((3l+r)T).
  \label{eq:xiLong}
\end{equation}
With a mapping similar to conventional CPM (Section \ref{sec:mapping})
we can simplify the two sums to a single term and obtain for $r=1,2$
\begin{equation}
  \xi_m(3l+r)=\frac{h}{2}d_{3l+r-\gamma+1}
  +c_{m,r}((3l+r)T)-c_{m,r+1}((3l+r)T).
  \label{eq:xi}
\end{equation}
As the data symbols are equal on each antenna, the difference between
two different $\xi_m(3l+r)$ does not depend on the data symbol
$d_{3l+r-\gamma+1}$. Thus, when choosing {\em parallel mapping}, $L^2$
orthogonality only depends on the correction factor.

To fulfill Eq. (\ref{eq:cond}) for all antennas we take
\begin{itemize}
\item for $m=1$, $m'=2$
  \begin{equation}
    a_r=\frac{2\pi}{3}=2\pi[c_{1,r}((3l+r)T)-c_{1,r+1}((3l+r)T)-c_{2,r}((3l+r)T)+c_{2,r+1}((3l+r)T)],
    \label{eq:ar11}
  \end{equation}
\item for $m=2$, $m'=3$
  \begin{equation}
    a_r=\frac{2\pi}{3}=2\pi[c_{2,r}((3l+r)T)-c_{2,r+1}((3l+r)T)-c_{3,r}((3l+r)T)+c_{3,r+1}((3l+r)T)]
    \label{eq:ar12}
  \end{equation}
\item and for $m=1$, $m'=3$ we consequently get
  \begin{equation}
    a_r=\frac{4\pi}{3}=2\pi[c_{1,r}((3l+r)T)-c_{1,r+1}((3l+r)T)-c_{3,r}((3l+r)T)+c_{3,r+1}((3l+r)T)].
    \label{eq:ar13}
  \end{equation}
\end{itemize}
The other three possible combinations of $m$ and $m'$ with $m\neq m'$
lead only to a change of sign and we get
$a_r=-2\pi/3, -2\pi/3,$ $-4\pi/3$, respectively. Due to the modulo $2\pi$
character of our condition, these are also valid solutions.

For simplicity, we assume similar correction factors for each time
slot $r$ of one Tx antenna $c_{m,1}(t)=c_{m,2}(t)=c_{m,3}(t)$. Since
Eq.  (\ref{eq:ar13}) arises from Eq. (\ref{eq:ar11}) and
(\ref{eq:ar12}), we have two equations and three parameters:
$c_{1,r}(t)$, $c_{2,r}(t)$ and $c_{3,r}(t)$. Hence we define
$c_{2,r}(t)=0$ for $r=1,2,3$ and we get
$c_{1,r}((3l+r)T)-c_{1,r+1}((3l+r)T)=1/3$ and
$c_{3,r}((3l+r)T)-c_{3,r+1}((3l+r)T)=-1/3$ for $r=1,2$. Codes
fulfilling these conditions form the family of Parallel Codes (PC). We
will now describe some possible solutions of this family.

An obvious solution for the correction factor is obtained for all
functions which are $0$ for $t=(3l+r)T$ and $\pm 1/3$ for
$t=(3l+r+1)T$, e.g.
\begin{equation}
  c_{1,r}(t)=-c_{3,r}(t)=\frac{2}{3}\cdot\frac{t-(3l+r)T}{2T}
\end{equation}
for $(3l+r)T\leq t\leq(3l+r+1)T$. We denote this solution as linear
parallel code (linPC). Of course, other choices are possible, e.g.
based on raised cosine (rcPC).

Another way of defining the correction factor is
\begin{equation}
  c_{1,r}(t)=-c_{3,r}(t)=\sum\limits_{i=1}^\gamma\frac{2}{3} q(t-i'T)
\end{equation}
for $(3l+r)T\leq t\leq(3l+r+1)T$. In that case we take advantage of
the natural structure of CPM, i.e. in Eq. (\ref{eq:xiLong}) all except
one summands cancel down, similar to the terms with the data symbols.
This definition has the advantage that we can merge the correction
factor and the data symbol in Eq. (\ref{eq:defphi}) and we obtain two
pseudo alphabets shifted by an offset (offPC) for the first and third
transmitting antenna

\begin{align}
  &\Omega_{d_1}=\left\{
    -M+1+\frac{2}{3h},-M+3+\frac{2}{3h},\ldots,M-1+\frac{2}{3h}\right\}\nn
  &\Omega_{d_3}=\left\{
    -M+1-\frac{2}{3h},-M+3-\frac{2}{3h},\ldots,M-1-\frac{2}{3h}\right\}\nonumber.
\end{align}

Consequently, this $L^2$-orthogonal design may be seen as three
conventional CPM signals with different alphabet sets $\Omega_d$,
$\Omega_{d_1}$and $\Omega_{d_3}$ for each antenna. In this method, the
constant phase offsets introduce frequency shifts. But as shown by the
simulations in next section, these shifts are quite moderate.

\subsection{Properties of PC CPM}
\label{sec:propert}
\subsubsection{Decoding}
\label{sec:decod}

The optimal receiver for the proposed codes relies on the computation
of a metric over complete ST blocks followed by a maximum-likelihood
sequence estimation (MLSE).

Here, the metric is evaluated using the $L^2$ norm
\begin{align}
  D_1= \int\limits_{3lT}^{(3l+3)T} \big|
  y_{1,r}(t)-\sum\limits_{m=1}^3 \alpha_{1,m}s_{m,r}(t) \big| ^2\D t.
  \label{eq:DISText}
\end{align}
For convenience, we use here only one receiving antenna but the
extension to more than one is straightforward. The distance in Eq.
(\ref{eq:DISText}) is obtained for all $pM^{\gamma+L_t-1}$ possible
variations of $s_{m,r}(t)$ corresponding to the paths of the trellis.

The number of states can be reduced in two ways.  First, by using the
orthogonality property of the proposed code, all cross-correlations in
Eq. (\ref{eq:DISText}) are canceled out and we obtain
\begin{align}
  D_2=\sum\limits_{m=0}^3
  \sum\limits_{r=0}^{3}\int\limits_{(3l+r)T}^{(3l+r+1)T}
  \big|y_{1,r}(t)-\alpha_{1,m} s_{m,r}(t)\big|^2\D t.
  \label{eq:DIST1}
\end{align}
We only have to consider $pM^\gamma$ paths for every $s_{m,r}(t)$. The
complexity of computing the distance is therewith $L_t^2pM^\gamma$
which corresponds to the necessary effort to decode three symbols of
three CPM signals.

Second, by taking advantage of the {\em parallel mapping} we are not
forced to decode block-wise. We can compute the distances symbol-wise
with
\begin{align}
  D_3=\int\limits_{\beta T}^{(\beta+1)T} \big|
  y_{1,r}(t)-\sum\limits_{m=1}^3 \alpha_{1,m}s_{m,r}(t) \big| ^2 \D t
  \label{eq:DIST2}
\end{align}
with $r=(\beta\mod L_T)+1$. Similarly to conventional CPM,
$s_{m,r}(t)$ has $pM^\gamma$ possible values which have to be
evaluated for every antenna $m$. By doing that, we reduce the paths of
the trellis but, at the same time, we increase the number of
transitions in the trellis.

The number of paths can be further reduced by using some special
properties of CPM. There exist numerous efficient algorithms for MLSE.
However, the efficiency of the detection algorithm is not in the
primary scope of this report and will be the subject of another
upcoming paper.

\subsubsection{Diversity}
\label{sec:diversity}

\begin{figure}
  \subfigure[offPC]{%
    \includegraphics[width=3.4in]{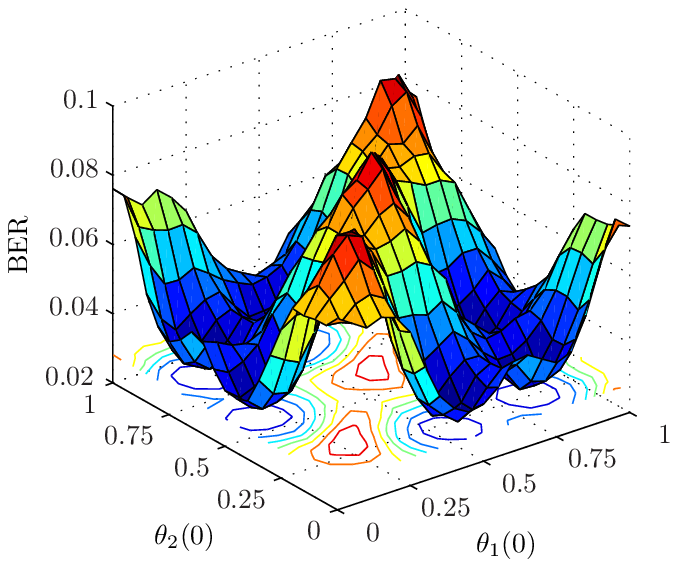}
    \label{fig:ab1BER}} \subfigure[linPC]{%
    \includegraphics[width=3.4in]{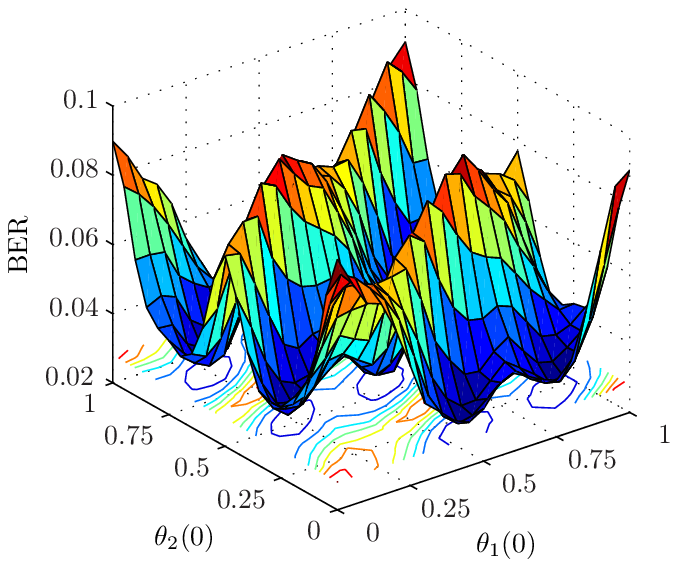}
    \label{fig:ab2BER}}
  \caption{Simulation results for the BER with different initial
    phases $\theta_i(0)$}
  \label{fig:abBER}
\end{figure}

Simulations of the proposed code show a similar behavior as codes with
full diversity. But, in contrast to $L^2$ orthogonal code for two
antennas \cite{Hess08}, the diversity of the three antenna code
depends on the initial phase $\theta_i(0)$ of each antenna $i$. Figure
\ref{fig:abBER} shows the simulation results for $\theta_3(0)=0$ and
varying $\theta_1(0)$ and $\theta_2(0)$. The bit error rate (BER)
clearly depends on the choice of the initial phase. This effect is
different for offset PC (Figure \ref{fig:ab1BER}) and linear PC
(Figure \ref{fig:ab2BER}).

The shown simulation results are using 4-array CPM ($M=4$) at
$E_b/N_0=13dB$. Further simulation with $M=8$ show no difference in
the location of the minimal BER. Whereas the varying modulation
indexes $h$ slightly change the position of the minima.

\subsection{Simulations}
\label{sec:simul}

\begin{figure}
 \begin{tabular}{cc}
  \includegraphics[width=3.4in]{BER} &
  \includegraphics[width=3.4in]{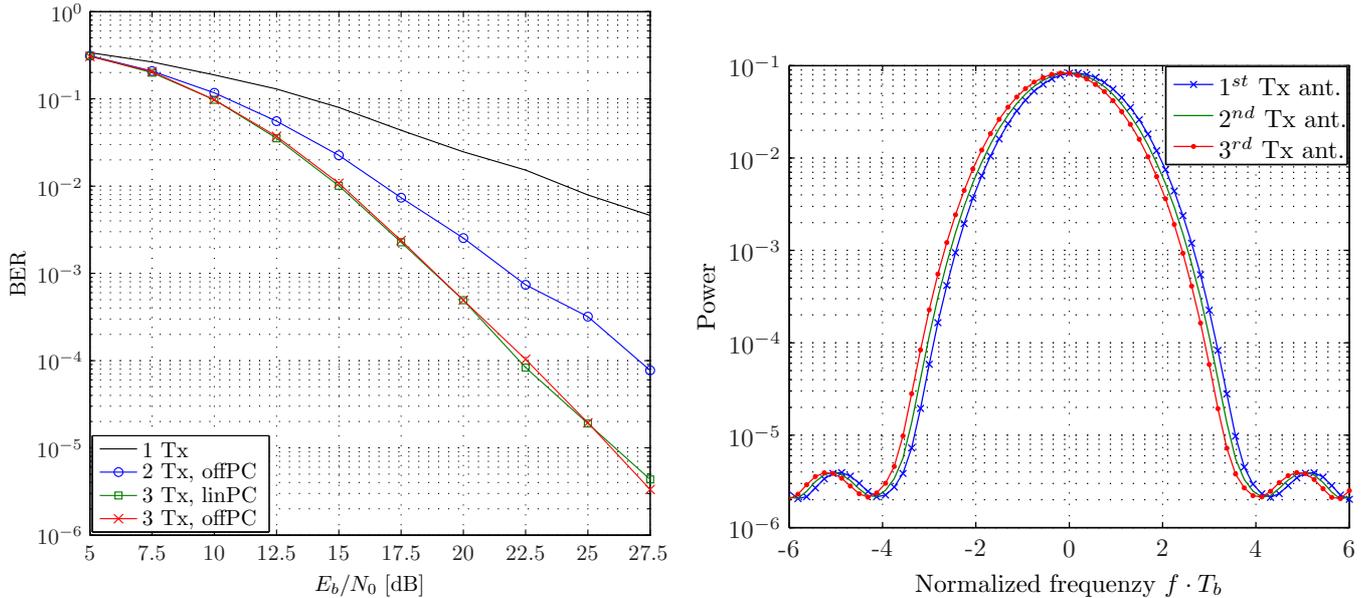}
\end{tabular}
  \caption{Left: Simulated bit error rate (BER) over a Rayleigh fading
    channel; Right: Power spectral density of the linPC analyzed with the Welch
    algorithm.}
  \label{fig:BER} \label{fig:PSS}
\end{figure}

In this section we test the proposed algorithms by running MATLAB
simulations. More precisely, we benchmark the offset parallel code
(offPC) for two and three Tx antennas and the linear parallel code
(linPC) for three Tx antennas. For all simulations we use a Gray-coded
CPM with a modulation index of $1/2$, an alphabet of 2 bits per symbol
$(M=4)$ and a memory length $\gamma$ of $2$. We use a linear phase
smoothing function $q(t)$ (2REC). Corresponding to section
\ref{sec:diversity}, we use $\theta_1(0)=0.75$, $\theta_2(0)=15$ and
$\theta_3(0)=0$ for linPC and $\theta_1(0)=0.45$, $\theta_2(0)=0.1$
and $\theta_3(0)=0$ for offPC.

The modulated signals are transmitted over a frequency flat Rayleigh
fading channel with complex additive white Gaussian noise. The fading
coefficients $\alpha_{n,m}$ are constant for the duration of a code
block (block fading) and known at the receiver (coherent detection).
To guarantee a fair treatment of single and multi antenna systems the
fading has to have a mean value of one.

The received signal $y_n(t)$ is demodulated by the methods introduced
in Section \ref{sec:decod}. Both require approximately the same
computational effort and achieve the same performance. The evaluated
distances are fed to the Viterbi algorithm (VA), which we use for
MLSE. In these demodulation methods, the trellis which is decoded by
the Viterbi algorithm has $pM^{\gamma-1}$ states and $M$ paths leaving
each state. In our simulations, the Viterbi algorithm is truncated to
a path memory of 10 code blocks, which means that we get a decoding
delay of $3\cdot 10T$.

Figure \ref{fig:BER} shows the simulations results for one, two and
three transmitting antennas. It can be seen that full diversity is
probably achieved and that linPC and offPC perform equal well if the
optimal initial phase is chosen.


A main reason for using CPM for STC is the spectral efficiency. Figure
\ref{fig:PSS} shows the simulated power spectral density (psd)
obtained by the analysis of $s_{m,r}(t)$ with the Welch algorithm. The
psd of the offPC has a negligible difference compared to the linPC.
Consequently it is not plotted in Figure \ref{fig:PSS}. The second Tx
antenna uses a conventional CPM signal without correction factor and
hence shows an equal psd. The spectra of the other antennas are
shifted due to the additional offset $c_{mr}(t)$ with a non zero mean.
Minimizing the $L^1$-norm of the difference between the unshifted and
shifted spectra result in a phase difference of $\pm 0.19$ measured in
normalized frequency $f\cdot T_d$, where $T_d=T/\log_2(M)$ is the bit
symbol length. Compared to the frequency offset of $0.375$ appearing
for two $L^2$-orthogonal antennas, the three antenna system requires
approximately the same bandwidth.

\subsection{Conclusion to part two}
\label{sec:concl}

In this part, we introduce a new family of $L^2$-orthogonal STC for
three antennas. These systems are based on CPM supplemented by
correction factors to ensure $L^2$-orthogonality. Structurally the
proposed code family has full rate and we expect full diversity.
Furthermore, we detail two simple representatives of the code family
(offPC, linPC), where the offPC offers better performance and a very
intuitive representation. By analyzing the power spectral density, it
is also shown, that the extension of the bandwidth, caused by the
correction factor, is small. Therefore the power efficiency of CPM is
maintained.

\section*{General conclusion}
In this report, we detail the construction and analyze the properties
of L2-orthogonal STC-CPM for two and three transmitting antennas. These codes
are attractive due to low-effort-decoding and the few restrictions the
code-family set upon parameters of CPM. Also, the simulation results
show the importance of optimizing the initial phases for an efficient
design and an optimal use of parallel codes.

\singlespacing \small \bibliographystyle{IEEEbib}

\bibliography{mh}

\begin{thebibliography}{10}

\bibitem{Zhan00}
X.~Zhang and M.~P. Fitz,
\newblock ``Space-time coding for {R}ayleigh fading channels in {CPM} system,''
\newblock {\em Proc. 38th Annu. Allerton Conf. Communication, Control, and
  Computing}, 2000.

\bibitem{Wang05}
D.~Wang, G.~Wang, and X.-G. Xia,
\newblock ``An orthogonal space-time coded partial response {CPM} system with
  fast decoding for two transmit antennas,''
\newblock {\em IEEE Trans. Wireless Commun.}, vol. 4, no. 5, pp. 2410 -- 2422,
  2005.

\bibitem{Alam98}
S.~M. Alamouti,
\newblock ``A simple transmit diversity technique for wireless
  communications,''
\newblock {\em IEEE J. Sel. Areas Commun.}, vol. 16, no. 8, pp. 1451 -- 1458,
  1998.

\bibitem{Taro99a}
V.~Tarokh, H.~Jafarkhani, and A.~R. Calderbank,
\newblock ``Space-time block codes from orthogonal designs,''
\newblock {\em IEEE Trans. Inf. Theory}, vol. 45, no. 5, pp. 1456 -- 1567,
  1999.

\bibitem{Zhan03}
X.~Zhang and M.~P. Fitz,
\newblock ``Space-time code design with continuous phase modulation,''
\newblock {\em IEEE J. Sel. Areas Commun.}, vol. 21, no. 5, pp. 783 -- 792,
  2003.

\bibitem{Zaji07}
A.~Zaji\'c and G.~St\"uber,
\newblock ``A space-time code design for partial-response {CPM}: Diversity
  order and coding gain,''
\newblock {\em IEEE ICC}, 2007.

\bibitem{Silv06}
A.-M. Silvester, L.~Lampe, and R.~Schober,
\newblock ``Diagonal space-time code design for continuous-phase modulation,''
\newblock {\em GLOBECOM}, 2006.

\bibitem{Wang04}
G.~Wang and X.-G. Xia,
\newblock ``An orthogonal space-time coded {CPM} system with fast decoding for
  two transmit antennas,''
\newblock {\em IEEE Trans. Inf. Theory}, vol. 50, no. 3, pp. 486 -- 493, 2004.

\bibitem{Ande86}
J.B. Anderson, T.~Aulin, and C.-E. Sundberg,
\newblock {\em Digital Phase Modulation},
\newblock Plenum Press, 1986.

\bibitem{Tela99}
I.~E. Telatar,
\newblock ``Capacity of multi-antenna gaussian channels,''
\newblock {\em European Trans. Telecommun.}, vol. 10, pp. 585 -- 595, 1999.

\bibitem{Fosc98}
G.~J. Foschini and M.~J. Gans,
\newblock ``On limits of wireless communications in a fading environment when
  using multiple antennas,''
\newblock {\em Wirel. Pers. Commun.}, vol. 6, no. 3, pp. 311--335, 1998.

\bibitem{Taro98}
V.~Tarokh, N.~Seshadri, and A.~R. Calderbank,
\newblock ``Space-time codes for high data rate wireless communication:
  Performance criterion and code construction,''
\newblock {\em IEEE Trans. Inf. Theory}, vol. 44, no. 2, pp. 744 -- 765, 1998.

\bibitem{Wang03}
G.~Wang, W.~Su, and X.-G. Xia,
\newblock ``Orthogonal-like space-time coded {CPM} system with fast decoding
  for three and four transmit antennas,''
\newblock {\em IEEE Globecom}, pp. 3321 -- 3325, 2003.

\bibitem{Hess08}
M.~Hesse, J.~Lebrun, and L.~Deneire,
\newblock ``L2 orthogonal space time code for continuouse phase modulation,''
\newblock in {\em {Proc.} IEEE ICC, {accepted for publication}}, 2008.

\end{thebibliography}

\end{document}